\documentclass{iucr}              % DO NOT DELETE THIS LINE

     \paperprodcode{a000000}      % Replace with production code if known
     \paperref{xx9999}            % Replace xx9999 with reference code if known
     \papertype{FA}               % Indicate type of article
                                       \paperlang{english}          % Can be english, french, german or russian
     \journalcode{A}              % Indicate the journal to which submitted
                                  
     \journalyr{2001}
     \journaliss{1}
     \journalvol{57}
     \journalfirstpage{000}
     \journallastpage{000}
     \journalreceived{0 XXXXXXX 0000}
     \journalaccepted{0 XXXXXXX 0000}
     \journalonline{0 XXXXXXX 0000}

\newcommand{\etal}{{\it et al\/}\ }

\begin{document}                  % DO NOT DELETE THIS LINE

     %-------------------------------------------------------------------------
     % The introductory (header) part of the paper
     %-------------------------------------------------------------------------

     % The title of the paper. Use \shorttitle to indicate an abbreviated title
     % for use in running heads (you will need to uncomment it).

\title{Solution of the crystallographic phase problem by iterated projections}
%\shorttitle{Solution of the phase problem}

\cauthor{Veit}{Elser}{ve10@cornell.edu}

\aff{Department of Physics, Cornell University, Ithaca, NY 14853-2501 \country{USA}}

%\shortauthor{Elser}
     
\maketitle                        % DO NOT DELETE THIS LINE

\begin{synopsis}
A new method of \emph{ab initio} phase determination is demonstrated which does not rely on 
Fourier-space formulations of atomicity.
\end{synopsis}

\begin{abstract}
An algorithm for determining crystal structures from diffraction data is described
which does not rely on the usual Fourier-space formulations of atomicity. The new algorithm
implements atomicity constraints in real-space, as well as intensity constraints in Fourier-space,
by projections which restore each constraint with the minimal modification of the scattering density. To recover
the true density, the two projections are combined into a single operation, the \emph{difference map}, 
which is iterated until the magnitude of the density modification becomes acceptably small. 
The resulting density, when acted
upon by a single additional operation, is by construction a density which satisfies both
intensity and atomicity constraints. Numerical experiments have yielded solutions for atomic
resolution x-ray data sets 
with over 400 non-hydrogen atoms, as well as for neutron data, where positivity of the density 
cannot be invoked.
\end{abstract}

\section{Introduction}

This year marks the semi-centennial of the realization by crystallographers 
that diffraction intensities
possess sufficient information to reconstruct an atomistic structure (Sayre, 2002). 
The simple fact that the scattering arises
from a known number of nearly point-like entities, while clearly not as intricate in content
as the body of collected intensities themselves, 
is by itself a significant piece of information. The first important steps
in utilizing atomicity in structure determination where taken by Sayre (1952) in his celebrated
equation, and later by Karle and Hauptman (1953) in their probabilistic analysis of structure factors. What is remarkable in the subsequent fifty-year history of direct methods, especially in view of the development
of the FFT already in the mid 1960s, is that atomicity has always been imposed in Fourier-space. The
efficiency of the transformation to real-space, made possible by the FFT, might have ushered in an
era where atomicity was imposed in the space where it is most naturally expressed. While
the most successful direct method programs, such as \texttt{SnB} (Miller \etal, 1994; Weeks \& Miller, 1999)
and \texttt{SHELXD} (Sheldrick, 1997, 1998), 
have adopted a significant degree of
atomicity intervention in real-space, the traditional Fourier-space approach to atomicity has continued to
be dominant in the development of algorithms.

The aim of the work detailed below was to develop a practical phase determination algorithm
for crystal structures that imposed atomicity entirely within real-space. A key component of the
algorithm is an iterative operation (difference map) that was discovered by deconstructing the most successful
algorithm (hybrid input-output) for the phase problem in optics (Fienup, 1982) 
and reexpressing it in terms having 
wider applicability (Elser, 2002a). Experiments with the \texttt{atom\_retriever} 
implementation of the new algorithm
on a variety of test structures demonstrate both its robustness and speed. The flexibility of the new
approach, with respect to the kinds of constraints that can be imposed in real-space, raises hopes of
an \textit{ab initio} solver not limited by atomic resolution data. 

\section{Constraints and projections}

The choice of the algorithm's fundamental variables is largely motivated by the mathematical structure
of the iterative step (Section \ref{diffMap}). In particular, the object that is iterated should have the property
that it can be added, in the sense of a linear vector space, to other objects, and that there is a natural
expression for the distance between objects. The unknown Fourier phases, for example, are not
good candidates in this respect. A better choice, and the one we adopt, is the real-space scattering 
density sampled on a finite regular grid. The relationship between real-space sampling and Fourier-space
sampling on the reciprocal lattice is quite direct, as illustrated by the two dimensional example in Figure 1.
Shown on the left (a) is the actual scattering density within one unit cell. The structure factors of the
corresponding crystal decay with scattering angle so that only a limited range about the origin in
reciprocal space are measured. By padding with zeroes at the corners, the Fourier-space measurements
can be fit into a finite rectangular grid as shown in the middle figure (b). Given phases for the structure
factors on the bounded Fourier-space grid, the discrete Fourier transform of the resulting complex
structure factors then gives the discretely sampled real-space
density shown on the left (c). Conversely, given a scattering density on the real-space grid (c), the
inverse Fourier transform gives the complex structure factors on the bounded Fourier-space grid, 
although with magnitudes not necessarily matching the measurements (b).

\subsection{Intensity constraints}\label{FourierProj}

A valid density in real-space must first of all have the property that the inverse Fourier transform
gives the measured structure factor \emph{magnitudes}. If not, one can seek the minimal density
modification that brings the actual magnitudes into agreement with the measured ones. Using the
symbol $\rho$ to represent the vector of densities on the real-space grid, the mathematical operation
which accomplishes this is the projection $\Pi_{\mathrm{F}}(\rho)$. The projected density is uniquely
defined by the properties that its Fourier transform has the correct (given) magnitudes and the distance
$\|\Pi_{\mathrm{F}}(\rho)-\rho\|$ is minimized. It is convenient to use the Euclidean distance since it
is preserved by Fourier transformation:
\begin{equation}\label{norm}
\|\rho\|^2 = {\textstyle \sum_{\mathbf{r}}}\rho_{\mathbf{r}}^2 
= {\textstyle \sum_{\mathbf{q}}}|\tilde{\rho}_{\mathbf{q}}|^2= \|\tilde{\rho}\|^2\; .
\end{equation}
In (\ref{norm}) the indices $\mathbf{r}$ and $\mathbf{q}$ denote grid points in real-space and 
Fourier-space (reciprocal lattice), respectively, and the complex structure factors
$\tilde{\rho}_{\mathbf{q}}$ are related to the real-space density by
\begin{equation}
\tilde{\rho}_{\mathbf{q}}=\left(\mathcal{F}(\rho)\right)_{\mathbf{q}}
=\frac{1}{\sqrt{M}}\sum_{\mathbf{r}}e^{2\pi \mathrm{i} 
\mathbf{q}\cdot\mathbf{r}}
\rho_{\mathbf{r}}\; ,
\end{equation}
where $M$ is the total number of grid points (in real or Fourier space). Using the unit cell's fractional coordinates
$(x,y,z)$ to label $\mathbf{r}$ and the Miller indices $(h,k,l)$ for $\mathbf{q}$, we have
\mbox{$\mathbf{q}\cdot\mathbf{r}=h x+k y+l z$}. The invariance of the distance (\ref{norm})
makes it possible to achieve the distance minimizing property of the projection $\Pi_{\mathrm{F}}$ very
easily in Fourier-space. Specifically, for every complex structure factor $\tilde{\rho}_{\mathbf{q}}$
we wish to find the nearest point in the complex plane lying on a circle corresponding to the measured
magnitude $F_{\mathbf{q}}$. The required projection is therefore accomplished by simply rescaling the
magnitude of the structure factor:
\begin{equation}\label{simpleFourierProj}
\left(\widetilde{\Pi}_{\mathrm{F}}(\tilde{\rho})\right)_{\mathbf{q}} = \frac{F_{\mathbf{q}}}{|\tilde{\rho}_{\mathbf{q}}|}
\tilde{\rho}_{\mathbf{q}}\; .
\end{equation}
A vanishing denominator in (\ref{simpleFourierProj}) is not a problem since it represents a set of measure
zero if one neglects extinctions (where $F_{\mathbf{q}}$ also vanishes). The projection in real-space is
expressed symbolically as:
\begin{equation}
\Pi_{\mathrm{F}} = \mathcal{F}^{-1}\,\widetilde{\Pi}_{\mathrm{F}}\,\mathcal{F}\; .
\end{equation}
When the Fourier transforms are implemented with the FFT, the computational cost of projecting a density
on the intensity constraints grows as $M\log{M}$.

A practical algorithm using the Fourier intensity projection $\Pi_{\mathrm{F}}$ must address the fact that
not all the structure factors on the Fourier-space grid will be measured. In addition to $F_{\mathbf{0}}$,
measurements near $\mathbf{q}=\mathbf{0}$ will frequently be absent or very unreliable, particularly
for large unit cell crystals. At the other extreme, structure factors in the corners of the Fourier-space grid
(Fig. 1(b)) will be absent because data is normally collected within an ellipsoidal domain about the
origin. The absent structure factors can be treated in a uniform way by applying \emph{bound} constraints rather
than value constraints. A bound $|\tilde{\rho}_{\mathbf{q}}|<F_{\mathbf{q}}^{\mathrm{B}}$
in Fourier-space is geometrically a disk, and the projection which restores the bound constraint either
leaves the structure factor unchanged, when it is inside the disk, or moves it to the nearest point on the
circumference, when it lies outside. Using $\mathcal{D}$ to denote the set of grid points $\mathbf{q}$ 
for which measured values $F_{\mathbf{q}}$ exist, and assuming bounds $F_{\mathbf{q}}^{\mathrm{B}}$
can be found for all others, equation (\ref{simpleFourierProj}) should be replaced by:
\begin{equation}
\left(\widetilde{\Pi}_{\mathrm{F}}(\tilde{\rho})\right)_{\mathbf{q}} = \cases{
(F_{\mathbf{q}}/|\tilde{\rho}_{\mathbf{q}}|)
\tilde{\rho}_{\mathbf{q}}& for $\mathbf{q}\in \mathcal{D}$\cr
(F_{\mathbf{q}}^{\mathrm{B}}/|\tilde{\rho}_{\mathbf{q}}|)
\tilde{\rho}_{\mathbf{q}}& for $\mathbf{q}\notin \mathcal{D}$ 
and $|\tilde{\rho}_{\mathbf{q}}|>F_{\mathbf{q}}^{\mathrm{B}}$\cr
\tilde{\rho}_{\mathbf{q}} & for $\mathbf{q}\notin \mathcal{D}$ and $|\tilde{\rho}_{\mathbf{q}}|<F_{\mathbf{q}}^{\mathrm{B}}$.}
\end{equation}
At large $\mathbf{q}$ one can obtain reliable bounds by extrapolating the measured structure factors
on a Wilson plot (see Sec. \ref{input}). Near the origin, where usually only few structure factors are absent, an infinite bound
is usually adequate and avoids a more difficult estimation problem. An example of a Fourier-space grid,
showing structure factor values and bounds taken from data for a 148-atom peptide structure
(Table 1, ref. 5), is shown in Figure 2.  

\subsection{Atomicity constraints}
Atomicity can be imposed as a support constraint, where the support $S$ of the density is a subset of the
grid points in real-space with the property $\rho_{\mathbf{r}}\neq 0$ implies $\mathbf{r}\in S$. 
However, in contrast
to phase retrieval with nonperiodic objects, 
where $S$ is known or can be bounded, in crystallography one only knows that $S$ has atomic
characteristics. The simplest definition of an \emph{atomic support} is the union of
a known number of compact subsets of grid points, each representing one atom, 
and having arbitrary locations within
the unit cell. Given a particular atomic support $S$, the projection $\Pi_S$ of an arbitrary density $\rho$,
to a minimally modified density having support $S$, is simply
\begin{equation}\label{supportProj}
\left(\Pi_S(\rho)\right)_{\mathbf{r}}=\cases{
\rho_{\mathbf{r}}& for $\mathbf{r}\in S$\cr
0& otherwise.}
\end{equation}
If $\mathcal{A}$ denotes the collection of all atomic supports (differing in atom locations), then atomicity
projection is defined by
\begin{equation}\label{atomProj}
\Pi_{\mathrm{A}}(\rho)=\Pi_{S^\prime}(\rho)\; ,
\end{equation}
where $S^\prime\in\mathcal{A}$ is the atomic support that minimizes $\|\Pi_S(\rho)-\rho\|$ over all
$S\in\mathcal{A}$. While (\ref{supportProj}) can be computed quickly, an exhaustive search over all
atomic supports in $\mathcal{A}$ to find (\ref{atomProj}) may be prohibitive. 
We therefore adopt a heuristic (described below) that quickly finds
an atomic support that is usually optimal.

In describing the precise projection operation we distinguish two cases: atomicity projection for positive atoms
(A+), and atomicity projection for atoms of arbitrary sign (A). The former is used with x-ray diffraction data, the
latter with neutron diffraction data when atomic species with both signs of scattering length are present. In
both cases we assume the number of atoms per unit cell $N$ is known. In the case of x-ray diffraction this
will usually not include H atoms.

The first step in computing the projection $\Pi_{\mathrm{A+}}(\rho)$ is to sort the density values on the real-space
grid. Then, beginning with the largest density, grid points with the 
property of being a local maximum are identified.
A local maximum is defined by having a larger density value than any of its 26 neighboring grid points. 
Each time a local
maximum is found, the 27 density values (maximum + neighbors) are copied, 
after \emph{positivity projection}, onto a 
real-space output grid that was
initially set to zero. Positivity projection, given by
\begin{equation}
\left(\Pi_{+}(\rho)\right)_{\mathbf{r}}=\cases{
\rho_{\mathbf{r}}& if $\rho_{\mathbf{r}}>0$\cr
0& otherwise,}
\end{equation}
is the minimal modification that restores the positivity of atoms.
The search through the sorted densities terminates 
when $N$ local maxima have been identified
and copied (positively) into the output grid. 
A graphical example of $\Pi_{\mathrm{A+}}$ in two dimensions (and 8 neighbors)
is shown in Figure 3(b).

Two modifications are required to compute the projection to atoms of arbitrary sign, $\Pi_{\mathrm{A}}(\rho)$.
To identify large peaks of arbitrary sign, the densities on the real-space grid are sorted by absolute
value. Then, densities in the sorted list are identified as local maxima or minima, depending on their sign.
Positivity projection is still applied to local maxima, whereas local minima are subjected to its counterpart,
negativity projection. Figure 3(c) shows the action of $\Pi_{\mathrm{A}}$. The computationally cost
of both types of atomicity projection is dominated by the sort of the $M$ densities on the real-space grid. Using
the quicksort algorithm this cost grows as $M\log{M}$, proportional to the cost of Fourier intensity projection
(Section \ref{FourierProj}).

The atomicity projections described require data with sufficient resolution. From the conventional
definition
\begin{equation}
(\mathrm{resolution})=d_{\mathrm{min}}=2\pi/Q_{\mathrm{max}}\; ,
\end{equation}
where $Q=(2\pi/\lambda)2\sin\theta$ is the magnitude of the physical scattering wavevector, one
can obtain in physical units the real-space grid spacing. The relationship between $Q$ and
the vector of Miller indices $\mathbf{q}$ is given by
\begin{equation}\label{Qtohkl}
(Q/ 2\pi)^2 = \mathbf{q}\cdot \mathbf{M}\cdot \mathbf{q}\to(h/a)^2+(k/b)^2+(l/c)^2\; ,
\end{equation}
where the matrix $\mathbf{M}$ is a metric constructed from the unit cell parameters, and the last expression
gives the explicit form for an orthorhombic cell with dimensions $a$, $b$ and $c$. For the ranges on the
Miller indices to be consistent with a $Q_{\mathrm{max}}$, equation (\ref{Qtohkl}) shows in particular that
\begin{equation}
|h|< a (Q_{\mathrm{max}}/ 2\pi)=a/d_{\mathrm{min}}\; .
\end{equation}
 The number of Fourier-space grid points for the index $h$ is therefore $2(a/d_{\mathrm{min}})$, and since the
real-space grid has the same number of points and has physical dimension $a$, the grid spacing is
$d_{\mathrm{min}}/2$.

According to our projection heuristic, a pair of local extrema
(of the same sign) can never be neighbors on the grid, but must be separated by at 
least two grid spacings in one of the
three dimensions. Supposing for simplicity that the unit cell is nearly orthorhombic, the most
problematic relative displacement for a pair of atom centers is along the body diagonal of the grid. In that
case the displacement must exceed $(2,2,2)$ in grid units, otherwise the corresponding local extrema
might be neighbors on the grid. The minimum separation of atomic centers must therefore
satisfy \mbox{$r_{\mathrm{min}}>2\sqrt{3}$} in grid units, or
\mbox{$r_{\mathrm{min}}>\sqrt{3}d_{\mathrm{min}}$}.
Since for organic structures \mbox{$r_{\mathrm{min}}\approx 1.4$\AA} $\,$(neglecting H atoms), this statement
implies \mbox{$d_{\mathrm{min}}<0.81$\AA}. On the other hand, this bound is derived 
from the worst-case placement (relative to the grid) of two atoms in a structure of many atoms. Atom pairs
displaced along a grid axis, for example, yield the more generous bound
\mbox{$d_{\mathrm{min}}<r_{\mathrm{min}}$}.
It is therefore not surprising that this form of atomicity projection has succeeded in solving organic
structures at resolutions exceeding $0.81$\AA $\,$(Table 1).

\section{The difference map}\label{diffMap}

Given two sets of constraints on the density, implemented respectively by projections $\Pi_1$ and $\Pi_2$, the
difference map is an iterative procedure for obtaining a solution density $\rho_{\mathrm{sol}}$ that
satisfies both constraints, specifically:
\begin{equation}\label{solution}
\Pi_1(\rho_{\mathrm{sol}})=\rho_{\mathrm{sol}}=\Pi_2(\rho_{\mathrm{sol}})\; .
\end{equation}
Although our main interest is the projections $\Pi_1=\Pi_{\mathrm{A}}$ (or $\Pi_{\mathrm{A+}}$) and
$\Pi_2=\Pi_{\mathrm{F}}$, we begin with a review of the solution method for a general pair of projections
(Elser, 2002a).

\subsection{Fixed points and solutions}

Starting with an arbitrary initial density $\rho(0)$, a sequence of iterates $\rho(n)=D^n\left(\rho(0)\right)$
is generated by repeated application of the difference map:
\begin{equation}\label{diffMapDefined}
D\colon \rho\mapsto\rho+\beta(\Pi_1\circ f_2-\Pi_2\circ f_1)(\rho)\; .
\end{equation}
Each of the projections in (\ref{diffMapDefined}) is composed
with a map
\begin{equation}\label{f_i}
f_i=(1+\gamma_i)\Pi_i-\gamma_i\quad (i=1,2) ;
\end{equation}
$\beta\neq 0$, $\gamma_1$ and $\gamma_2$ are real parameters.
The difference map has two key properties, the first being that a solution, as defined by
(\ref{solution}), exists if and only if the map has a fixed point, $\rho^*=D(\rho^*)$. 
To see this, note that the
difference in (\ref{diffMapDefined}) vanishes at a fixed point, hence:
\begin{equation}\label{rhoPrime}
\Pi_1\circ f_2(\rho^*)=\Pi_2\circ f_1(\rho^*)=\rho^\prime\; .
\end{equation}
Applying either of the projections to (\ref{rhoPrime}) and using the property $\Pi_i\circ\Pi_i=\Pi_i$,
we obtain
\begin{equation}
\Pi_1(\rho^\prime)=\rho^\prime=\Pi_2(\rho^\prime)\; ,
\end{equation}
thus identifying $\rho^\prime$ with $\rho_{\mathrm{sol}}$. Conversely, if $\rho_{\mathrm{sol}}$ exists,
the set of fixed points is nonempty since it is easily verified that 
\mbox{$D(\rho_{\mathrm{sol}})=\rho_{\mathrm{sol}}$}.

The fixed point property of the difference map makes no reference to the detailed forms of the maps $f_i$.
These maps are key to the second property: the attractive nature of the fixed points. In an iterative solution
method a fixed point is useless unless it is attractive; moreover, the greater the basin of attraction, the more
effective is the method. Replacing the $f_i$ by identity maps, for example, creates unstable (repulsive) directions
in a fixed point's local behavior (Elser, 2002a), effectively reducing to zero the probability of arriving at the fixed point. The chosen form (\ref{f_i}) of the maps $f_i$ is the simplest, involving just the projections, 
that for suitable values
of the parameters $\gamma_i$ renders the fixed points of the difference map attractive.

Fixed points of the
difference map should not be confused with the solution. 
The former are not unique, comprising in fact a submanifold
in the space of densities; formally, the \mbox{submanifold} of fixed points is given by the 
intersection of inverse images:
\begin{equation}
(\Pi_1\circ f_2)^{-1}(\rho_{\mathrm{sol}})\cap (\Pi_2\circ f_1)^{-1}(\rho_{\mathrm{sol}})\; .
\end{equation}
During the course of iterating the difference map the convergence to a fixed point is assessed by the norm
of the difference:
\begin{equation}\label{epsilon}
\epsilon_n = \|(\Pi_1\circ f_2-\Pi_2\circ f_1)\left(\rho(n)\right)\|\; .
\end{equation}
When $\epsilon_n$ becomes acceptably small, $\rho_{\mathrm{sol}}$ is obtained, as in (\ref{rhoPrime}),
by applying $\Pi_1\circ f_2$ (or $\Pi_2\circ f_1$) to the estimate $\rho^*\approx\rho(n)$.

\subsection{Parameter values}

For a particular $\beta$, interpreted as a step size, the values of $\gamma_1$ and $\gamma_2$ are selected
to optimize the convergence at fixed points. The earliest analysis of the difference map (Elser, 2002a) assumed
local orthogonality of the two constraint subspaces and found
\begin{eqnarray}
\gamma_1&=&-1/\beta\\
\gamma_2&=&1/\beta\; .\label{orthoCase}
\end{eqnarray}
Subsequent work (Elser, 2002b) considered an average-case analysis for particular kinds of constraints, including
atomicity, and found optimal parameters \mbox{($1\to \mathrm{S}$, $2\to \mathrm{F}$)}
\begin{eqnarray}
\gamma_{\mathrm{S}}&=&-1/\beta\\
\gamma_{\mathrm{F}}&=&\frac{1+t_{\mathrm{F}}(1-\beta)}{\beta}\; ,\label{aveCase}
\end{eqnarray}
where $t_{\mathrm{F}}\approx 0.5$ is the fraction of 
Fourier-space grid points with known (non-negligible) structure factors,
and the projection $S$ (support) corresponds to atomicity (A or A+). However, due to the fact that the
typical step size is $\beta\approx 0.7$, the small numerical difference between 
(\ref{orthoCase}) and (\ref{aveCase})
has not led to a significant change in performance of the algorithm. All experiments quoted in this work used
the simpler expression (\ref{orthoCase}).

Since there is as yet no theory for determining the optimal value of $\beta$ for any particular application, $\beta$
remains the single parameter of the algorithm that must be optimized empirically. Average solution times,
measured in terms of difference map iterations, are shown in Figure 4 as a function of $\beta$ for a 148-atom
peptide structure (Table 1, ref. 5). Systematic trends in optimal $\beta$ values with data resolution and
$M/N=(\mathrm{grid size})/(\mathrm{atoms})$ have not been performed, although it appears
that optimal values fall in the range shown ($0.4\le\beta\le 0.8$).

Some algorithms (\texttt{SnB}, \texttt{SHELXD}) treat the number of non-H atoms as a parameter,
with $N$ deliberately chosen significantly smaller than the best estimate of the 
actual number to improve performance.
Preliminary studies with the difference map algorithm on the 148-atom peptide structure showed only
a very weak variation in the average number of iterations when $N$ was varied by
$\pm 8\%$. This result indicates that while $N$ is not a useful parameter, the algorithm can
tolerate inevitable uncertainties in the actual numbers of atoms.
From a logical viewpoint, choosing $N$ \emph{larger} than the actual number of atoms is
still valid: the atomicity constraint has simply been weakened.

\subsection{Convergence with imperfect data}

When formulated in terms of constraints, the uniqueness of solutions to the phase problem requires an
overconstrained situation. Considered geometrically, the two sets of constraints (say intensity and atomicity)
are individually submanifolds in the space of densities having relatively low dimensionality. Specifically,
in the overconstrained case the sum of the dimensionalities is less than that of the ambient space ($M$),
such that the intersection of generic submanifolds, of the same dimensions, 
would be empty (rather than a submanifold of positive dimension).
The constraint submanifolds in a well-posed phase problem, given perfect data, are 
nongeneric in the sense
that a solution is known to exist, or equivalently, the submanifolds have a nonempty intersection
in spite of their low dimensionality. The fine-tuning
implicit in the intensity data (say) required to achieve an intersection, or true solution, is upset by practically
any departures from ideality. Chief among these in the crystallographic phase problem are statistical errors
in the intensity measurements and the neglect of hydrogen atoms in the treatment of atomicity. Faced with
these realities, one must abandon the hope of finding a solution in the strict sense.

Although the constraint submanifolds are not expected to perfectly intersect with realistic
noisy data, we expect them to have a small separation (in the space of densities) 
in the vicinity of the true density. The convergence
estimate $\epsilon$ (see (\ref{epsilon})) can be interpreted as the currently achieved distance between
constraint submanifolds, and solutions should be identified not by its vanishing but by its value dropping a
significant amount. Plots of $\epsilon_n$ as a function of iteration $n$ are contrasted in Figure 5, 
for synthetically generated (top)
and experimental (bottom) data. Experimental data for the 148-atom peptide structure (Table 1, ref. 5) 
was used for the
imperfect data set, and synthetic data for a 148-equal-atom structure having the same $M/N$ ratio was used
to simulate perfect data. To ensure perfect compliance with the atomicity constraint, the Gaussian atoms used
to create the perfect data where given supports on $3\times 3\times 3$ subsets of the real-space grid; atom centers,
given by a random number generator, avoided the minimum separation 
$r_{\mathrm{min}}=\sqrt{12}$ (grid units). The sharp drop in $\epsilon$ 
displayed by the experimental data, and
observed in all difference map solutions reported here, demonstrates the viability of $\epsilon$ as a solution criterion
even in the case of imperfect data.

With imperfect data it appears that the minimum $\epsilon$ occurs shortly after the initial drop, and is not
surpassed subsequently. As a best estimate of a fixed point density we therefore take the difference map iterate
at the $\epsilon$ minimum; the corresponding solution estimate is found by applying 
$\Pi_{\mathrm{F}}\circ f_{\mathrm{A+}}$ or ($\Pi_{\mathrm{F}}\circ f_{\mathrm{A}}$). Using this procedure
on the known 148-atom peptide structure gave a mean figure-of-merit $\langle\cos{\Delta\phi}\rangle=0.71$
when averaged over all reflections in the data.

The overconstrained nature of the problem solved by the difference map can be appreciated by closer examination
of the solution found with perfect data. Shortly after the sharp drop in $\epsilon$ when the solution is first found,
$\epsilon$ decays monotonically to zero. This behavior implies that the structure factors of the density 
at large angles, that
were provided only as bounds to the algorithm, are in fact being extrapolated to their true values by the
iterative process (as was confirmed by direct examination of the solution's structure factors).

\subsection{The solution process}

The problem of finding a point in Euclidean space that satisfies a number of constraints, 
or showing that no such point exists, is known as a \emph{feasibility problem} in the optimization literature.
Theoretical studies have mostly focused on the case of convex constraint subspaces, where monotonicity
of convergence can usually be proven for a variety of iterative methods. 
However, since both sets of constraints in the crystallographic phase
problem (Fourier intensity and atomicity) are \emph{nonconvex}, no rigorous results are available. The local
analysis of the difference map, quoted above, only establishes the favorable fixed point characteristics 
of the map, and provides no estimate on the number of iterations required to enter a fixed point's sphere
of influence.

A dynamical systems perspective, combined with empirical data, provides a useful, though
nonrigorous, picture for the difference map's mode of operation. Some salient features of the evolution
of the density are illustrated in Figure 6. Relatively rapid changes occur on the scale of very few iterations, and
continue in an apparent steady-state until quite abruptly the fixed point is encountered. During the long period
of rapid changes there is no obvious progress toward the solution and the dynamics is
well characterized as chaotic in the \emph{strongly mixing} regime, where iterates settle into a stationary probability
distribution very quickly. The basin of attraction of the map's fixed points has some overlap with this
probability distribution, the magnitude of which determines the mean number of iterations required to arrive at the
solution when averaged over starting points. 

Taking this interpretation as a hypothesis, it can be tested by
compiling a distribution of solution times for a given problem instance, as measured by the number of iterations
before the sharp drop in $\epsilon$ occurs (Fig. 5). If the strongly mixing property holds, then iterates are
effectively subject to a probability of arriving at the attractive basin of a fixed point that is constant in time, 
and hence solution times
will have an exponential distribution. An experiment comprising 3800 trials for the 148-atom peptide structure 
(all with $\beta=0.7$)
showed exactly this distribution. A solution was found in each trial, the longest requiring 7760 iterations, and the
mean for all trials was 1100 iterations. The distribution of iterations, normalized relative to the mean, is plotted in
Figure 7 and compared with the exponential distribution. Overall the agreement is very good: the slight deviation
at small iterations can be explained by a combination of the fixed (but small) number of iterations required 
to converge, first
to the stationary probability distribution, then, after arriving at the attractive basin, to the
fixed point.

The observed distribution of solution times greatly simplifies the solution protocol and eliminates yet another
potential parameter: the bound on the number of iterations. Iteration bounds, typically some multiple of the 
number of atoms $N$, are imposed by \texttt{SnB} and \texttt{SHELXD} and results are quoted in terms
of ``success rates". Given an exponential distribution of solution times (Fig. 7),
such a bound for the difference map algorithm is arbitrary since it will have no effect on the number 
of solutions found per total iterations 
performed on all trials. Expressed in more direct terms: the performance of the algorithm is practically unaffected
by random restarts, and hence there is no degradation of performance when iterations are allowed to continue
indefinitely. It is possible, however, that this conclusion will have to be modified if further experimentation with
the difference map, say with small $\beta$, finds a nonexponential distribution of solution times.

\section{Studies of test structures}

\subsection{The \texttt{atom\_retriever} computer program}

A preliminary implementation of the difference map algorithm for crystallographic applications exists as
the C-language program \texttt{atom\_retriever}. The software in its current form is best characterized
as a library of general-purpose subroutines that manipulate data in the uniform format of discretely sampled
densities. At the lowest level are subroutines for performing a variety of projections,
Fourier intensity and atomicity projection being of primary interest to crystallographers. The next level
of subroutines combine the chosen projections into the difference map. Finally, at the highest level are a
collection of drivers and translators, that provide options for monitoring and terminating iterations, as well
as converting input structure factor data files into the rectangular arrays used by the algorithm. With this
degree of transparency, it is hoped that users will experiment with innovations at the level of the projections,
which are at the heart of the algorithm's success.

\subsection{Space groups and structure factor input}\label{input}

The primary input to \texttt{atom\_retriever} is the rectangular array, containing structure factor values
and bounds (Fig. 2), and used by the Fourier intensity projection subroutine. Software for applying
symmetry elements to structure factor data in the construction of these arrays is still being
developed, limiting applications to structures with triclinic (P1 and P$\overline{1}$) space groups. 
The complete
set of space groups can be implemented by preparing the initial density $\rho(0)$ with a projection that
recognizes in Fourier space the phase relationships between structure factors for the specified group; no
changes are necessary in the 
\texttt{atom\_retriever} program, since, apart from numerical rounding
effects, the difference map preserves the symmetry of the density. Since symmetry projection
software is also under development, the P$\overline{1}$ structures studied to date were treated as P1, with
twice the true number of symmetry inequivalent atoms.

The truncation of the individual atomic supports during atomicity projection to a $3\times 3\times 3$ array of grid
points was shown by Elser (2002a) to be optimal when the corresponding Gaussian atom 
has a mean square displacement $\langle u_x^2\rangle \approx 0.55$ in grid units, or
\begin{equation}
B/8\pi^2=\langle u_x^2\rangle \approx 0.14\, d_{\mathrm{min}}^2\; ,
\end{equation}
where $B$ is an effective isotropic temperature factor. Defining a dimensionless temperature factor by
\begin{equation}\label{b}
b=B/d_{\mathrm{min}}^2\; ,
\end{equation}
one finds that typical x-ray and neutron data sets (see Table 1) satisfy $b<b_{\mathrm{opt}}\approx 11$. 
This means that $3\times 3\times 3$ is
usually a generous support, perhaps even sufficient to accommodate hydrogen neighbors.  

An effective temperature factor $B$, which combines the effects of atomic size, thermal
vibration and certain kinds of static disorder, is estimated from the data
by making a linear least-squares fit of pairs
$\{Q^2, \log{|F_Q|^2}\}$ to the form
\begin{equation}\label{fit}
\log{|F_Q|^2}=A-\frac{1}{2}B (Q/2\pi)^2\; ,
\end{equation}
for measurements in a restricted range $Q>Q_0$, where typically $2\pi/Q_0=1.2$\AA.
At large spatial frequencies the structure factors are well modeled as complex Gaussian random variables
(isotropic with mean 0) and hence the distribution of intensities at wave vector $Q$ is given by
\begin{equation}\label{intensityPDF}
P_Q(I)\,dI = \exp{\left(-I/\langle I_Q\rangle\right)}\, dI/\langle I_Q\rangle\; .
\end{equation}
Since a least-squares fit applied to (\ref{fit}) gives a formula for the average of $\log{I_Q}$, the distribution
(\ref{intensityPDF}) gives the result
\begin{equation}\label{logI}
A-\frac{1}{2}B (Q/2\pi)^2 = \langle \log{I_Q}\rangle = \log{\langle I_Q\rangle}-\gamma\; ,
\end{equation}
where $\gamma$ is Euler's constant. 

The probability distribution (\ref{intensityPDF}) is also used in the
determination of bounds on the magnitudes of unmeasured structure factors with $Q>Q_0$, the bulk being in the
corners of the Fourier-space grid where $Q>2\pi/d_{\mathrm{min}}$. 
If $M^\prime$ is the number of such (symmetry inequivalent) structure factors, then
$1/M^\prime$ is an acceptable probability for an actual intensity to exceed the bound. Equating this
with the probability $I>(F_Q^{\mathrm{B}})^2$ computed using (\ref{intensityPDF}), gives
\begin{equation}
F_Q^{\mathrm{B}} = \sqrt{\langle I_Q\rangle \log{M^\prime}}\; ,
\end{equation}
where $\langle I_Q\rangle$ is given explicitly in terms of the parameters $A$ and $B$ of the fit by
(\ref{logI}). There are far fewer unmeasured structure factors with $Q<Q_0$, and the
bound $F_Q^{\mathrm{B}} = \infty$ was used for these in all the studies reported here.

\subsection{Discussion of tests}\label{tests}

The results of a selection of tests of the \texttt{atom\_retriever} program are summarized in Table 1.
In each test the entire available set of experimental structure factors was used, with missing data replaced by
bounds in the manner described above; the grid dimensions correspond to $(2|h|_{\mathrm{max}}+2)\times
(2 |k|_{\mathrm{max}}+2)\times(2 |l|_{\mathrm{max}}+2)$. No effort was made to individually optimize 
performance with respect to $\beta$; the value chosen, $\beta = 0.7$, is near the minimum in
the average number of
iterations for the 148-atom peptide structure (see Fig. 4). Mean figures of merit $\langle\cos{\Delta\phi}\rangle$
for the P1 structures were determined relative to calculated phases. For the P$\overline{1}$ 
structures, where calculated
phases were not available, an internal figure of merit was determined by treating the structures as P1 (with
twice as many atoms) and taking the nearest set of centrosymmetric phases as the true phases.
When the number of iterations required by the algorithm has a broad distribution
(see Fig. 7),
the average number is quoted.

Atomicity projection for positive atoms ($\Pi_{\mathrm{A+}}$) was used
in all the x-ray data sets; the neutron data set for the mineral montebrasite
provided the sole application of the projection to atoms of arbitrary sign, $\Pi_{\mathrm{A}}$.
Nuclei with negative scattering length, such as Li, show up as light contrast in a field of dark
atoms in plots of the scattering density $\rho_{\mathrm{sol}}$. Figure 8 shows the scattering
density in one layer
of the montebrasite solution. The small number of iterations required to find the solution is typical
of few-atom structures, where the error diagnostic (\ref{epsilon}) decreases almost monotonically 
with iteration (Fig. 8). When appropriately translated, the scattering density was nearly centrosymmetric,
the resulting internal figure of merit being significantly larger than what would be obtained from
random phases: $\langle\cos{\Delta\phi}\rangle_{\mathrm{rand}}=2/\pi\approx 0.64$.

Rapid solutions with a near monotonic error decrease were also observed for the other few-atom structures:  
nitramine, pyrrole, punctaporonin
and triphenylphosphine. 
The nitramine and pyrrole
structures were selected for their high density and cell aspect ratios respectively. These characteristics had
no noticeable effects on the algorithm's performance. Triphenylphosphine, interestingly, 
has more atoms (when treated as P1) than the 148-atom
peptide structure which requires many more iterations. Data resolution is not the cause of this anomaly, as was
confirmed by truncating the triphenylphosphine data to the same 0.96 resolution as the peptide: the solution
was again found in only 25 iterations. A more likely cause is the presence of a moderately heavy
P atom in each of the eight triphenylphosphine molecules of the structure, in contrast to the near
equality of the non-H atoms
of the peptide. Arbitrary atomic charges are trivially accommodated by
the form of atomicity projection used by \texttt{atom\_retriever}. A comparably non-specific
atomicity was achieved relatively late in the development of the Fourier-space based framework 
(Hauptman, 1976; Rothbauer, 2000). 

It is premature to assess the algorithm's prospects in solving large structures from atomic resolution data.
The average solution time for the 148-atom peptide was 35 seconds on a 2GHz Pentium 4 (single processor),
not far behind \texttt{SnB2.2} (30 seconds) and \texttt{SHELXD} (11 seconds). More critical in evaluating performance is the growth in the average number of iterations with structure size. 
The largest structure attempted,
a synthetic $\alpha$-helical bundle, required nearly a half-million iterations per solution (about 30 hours).
Figure 9 shows the error estimate for a typical run, together with electron density contours obtained
directly from the discretely sampled density $\rho_{\mathrm{sol}}$.
The single, noticeably larger peak in the density was
identified with the chloride ion in the structure. This synthetic peptide differs from the 
smaller structures
in that the number of ordered, non-H atomic sites ($N$) is not known \textit{a priori} because of
the solvent contribution. The results quoted all
used $N=479$, the number of non-H atoms discovered in the original refinement 
(Priv\'{e} \etal, 1999), but the structure
was also solved with $N$ as small as 440. 

\section{Conclusions}

The crystallographic phase problem consists of two, logically distinct, though technically coupled, parts.
Using the terms ``needles" for solutions and ``hay" for non-solutions, these parts are: (1) distinguishing needles from hay, and (2) finding the proverbial needle in a haystack. As this work hopefully demonstrates, it is quite
straightforward to recognize needles when presented with one: one only requires projections that act trivially
(with negligible change) when operating on an object (\textit{i.e.} scattering density) that satisfies all the known
constraints (Fourier intensity, atomicity). If the constraints are too weak, as in a low resolution data set, the
phase problem may not be soluble in principle, because needles cannot be distinguished from the hay. 
On the other hand, even for low resolution data it is known (Podjarny \etal, 1987) 
that in certain circumstances needles can
still be recognized. A properly phased protein crystal, for example, will often have a well defined solvent
region and a characteristic histogram of density values (possibly at multiple length scales) within
the body of the molecule, even when individual atoms are not resolved. If a projection
(minimal density modification) can be constructed
that applies in this more general setting, the first part of the phase problem can be said to be solved.

The difference map solves the second part of the phase problem by providing
a uniform scheme for combining the applicable projections into an algorithm
that finds the needle. Its eventual success is practically guaranteed if the corresponding 
constraints are strong enough
to make needles distinct from hay. Perhaps most remarkable of all is the empirical fact 
that needles can be found in a reasonable time at all. Though the attractive basins of the difference map's
fixed points are, by proper choice of parameters, tuned to be as large as possible
(Elser, 2002a, 2002b), this local
optimization cannot predict the average solution time. With more experience we
anticipate a body of empirical relationships between average solution times and characteristics of the data
that can fill this gap.
Some progress in this direction 
was recently achieved (Elser, 2002c) in a highly idealized version of the phase problem (Zwick \etal, 1996), 
where the discretely sampled (one dimensional) density is known to be two-valued (a binary sequence).

     % Appendices appear after the main body of the text. They are prefixed by
     % a single \appendix declaration, and are then structured just like the
     % body text.

%\appendix
%\section{Appendix title}

     %-------------------------------------------------------------------------
     % The back matter of the paper - acknowledgements and references
     %-------------------------------------------------------------------------

     % Acknowledgements come after the appendices

\ack{Acknowledgements}

\texttt{atom\_retriever} was written while the author was a guest of the Center
for Experimental and Constructive Mathematics at Simon Fraser University. Encouragement and
technical suggestions from the director, Jonathan Borwein, and staff, in particular Rob Ballantyne,
are gratefully acknowledged. \texttt{atom\_retriever} would not have made the transition to a 
practical piece of software without George Sheldrick's generosity in sending progressively more
challenging data sets along with \texttt{SHELXD} benchmarks. Charles Miller kindly provided the \texttt{SnB}
benchmark quoted in Section \ref{tests}. The free use of experimental data sets is greatly
appreciated, with particular thanks to Bryan Chakoumakos (Table 1, ref.1) and George Sheldrick 
(Table 1, refs. 4 \& 5). Nick Elser was instrumental in adapting \texttt{atom\_retriever} 
to run on the Cornell Theory Center computing cluster. 
This research was conducted using the resources of the Cornell Theory Center, 
which receives funding from Cornell University, New York State, federal agencies, foundations, and corporate partners. The author's primary source of support is the National Science Foundation under grant ITR-0081775.

     % References are at the end of the document, between \begin{references}
     % and \end{references} tags. Each reference is in a \reference entry.

     %-------------------------------------------------------------------------
     % TABLES AND FIGURES SHOULD BE INSERTED AFTER THE MAIN BODY OF THE TEXT
     %-------------------------------------------------------------------------

     % Simple tables should use the tabular environment according to this
     % model

\newpage
\onecolumn

\begin{table}
\caption{Test structures solved by \texttt{atom\_retriever}.}
\begin{tabular*}{8in}{lcclllcc||cclc}      % Alignment for each cell: l=left, c=center, r=right
 structure    & ref.    & data       & refl's	& $N$       & group  & $d_{\mathrm{min}}$ & $b$
& grid & $\beta$ & iter's &  $\langle\cos{\Delta\phi}\rangle$  \\
\hline
montebrasite      & 1      & neutron      & 1024     & 20      & P1$^*$     & 0.66      & 2.5     & $22\times 24\times 22$      & 0.7	& 20      & 0.76\\

TEX nitramine	& 2	& x-ray	& 1911	& 36	& P1$^*$	& 0.76	& 6.6	& $20\times 22\times 24$	& 0.7	& 15	& 0.77\\

amphiphilic pyrrole	& 3	& x-ray	& 4238	& 48	& P1$^*$	& 0.81	& 11.0	& $14\times 26\times 52$	& 0.7	& 25	& 0.76\\

punctaporonin D      & 4      & x-ray      & 3764     & 54      & P1      & 0.85      & 10.4     & $22\times 26\times 30$      & 0.7	& 75      & 0.72\\

$\alpha$-helix \& $3_{10}/\alpha$-helix & 5 & x-ray & 7489
& 148     & P1      & 0.96      & 7.5	& $22\times 36\times 42$      & 0.7     & 1100      & 0.71\\

triphenylphosphine      & 6      & x-ray      & 9982     & 152      & P1$^*$     & 0.84      & 2.9     & $28\times 36\times 42$     & 0.7	& 25      & 0.79\\

$\alpha$-helical bundle & 7 & x-ray 	& 23681
& 479     & P1      & 0.90      & 10.1	& $40\times 46\times 60$ & 0.7     & 450000      & 0.75\\

\end{tabular*}
\hline
$^*$ The actual structure has symmetry P$\overline{1}$, but was treated as P1 with twice as many atoms.
\begin{enumerate}
\item Groat, L. A., Chakoumakos, B. C., Hoffman, C. M., Morell, H., Fyfe, C. A. \& Schultz, A. J. (2002).
\emph{American Mineralogist}, in press.
\item Karaghiosoff, K., Klap\"{o}tke, T. M., Michailovski, A. \& Holl, G. (2002). \emph{Acta. Cryst.} \textbf{C58}, 580-581.
\item Silva, M. R., Beja, A. M., Paix\~{a}o, J. A., Sobral, A. J. F. N., Lopes, S. H. \& Gonsalves, A. M. d'A. R. (2002). \emph{Acta. Cryst.} \textbf{C58}, 572-574.
\item Poyser, J. P., Edwards, R. L., Anderson, J. R., Hursthouse, M. B., Walker,
N. P. C., Sheldrick, G. M. \& Whalley, A. J. S. (1986). \emph{J. Antibiotics} \textbf{39},
167-169.
\item Karle, I. L, Flippen-Anderson, J. L., Uma, K., Balaram, H. \& Balaram, P.
(1989). \emph{Proc. Natl. Acad. Sci. USA} \textbf{86}, 765-769.
\item Kooijman, H., Spek, A. L., van Bommel, K. J. C., Verboom, W. \& Reinhoudt, D. N. (1998). 
\emph{Acta Cryst.} \textbf{C54}, 1695-1698.
\item Priv\'{e}, G. G., Anderson, D. H., Wesson, L., Cascio, D. \& Eisenberg, D. (1999).
\emph{Protein Science} \textbf{8}, 1400-1409.
\end{enumerate}
\end{table}

     % Postscript figures can be included with multiple figure blocks
\newpage

\begin{figure}
\caption{Discrete sampling of a scattering density in two dimensions. 
(a) Density within one unit cell. 
(b) Structure factor magnitudes for the density in (a), sampled over a range with $h^2 + k^2 < 10^2$.
(c) Discrete sampling of (a) obtained by applying the discrete Fourier transform to the \emph{complex} structure
factors (Fig. (b) combined with phases).}
\includegraphics{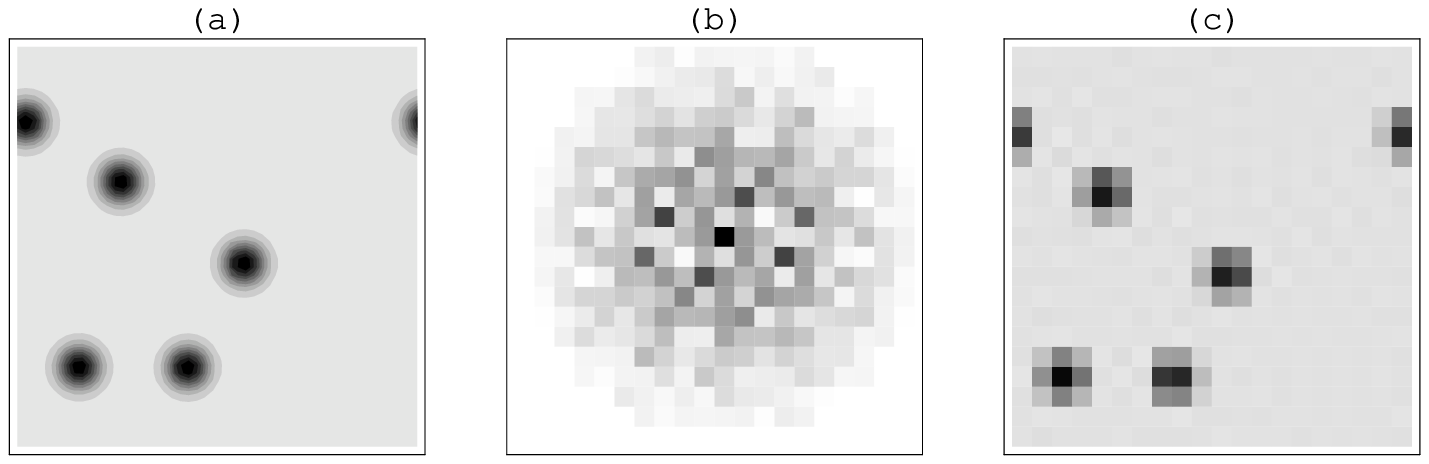}
\end{figure}

\begin{figure}
\caption{Structure factors in the plane $h=0$ for a 148-atom peptide 
structure (Table 1, ref. 5): (a) measured values; (b) bounds.}
\includegraphics{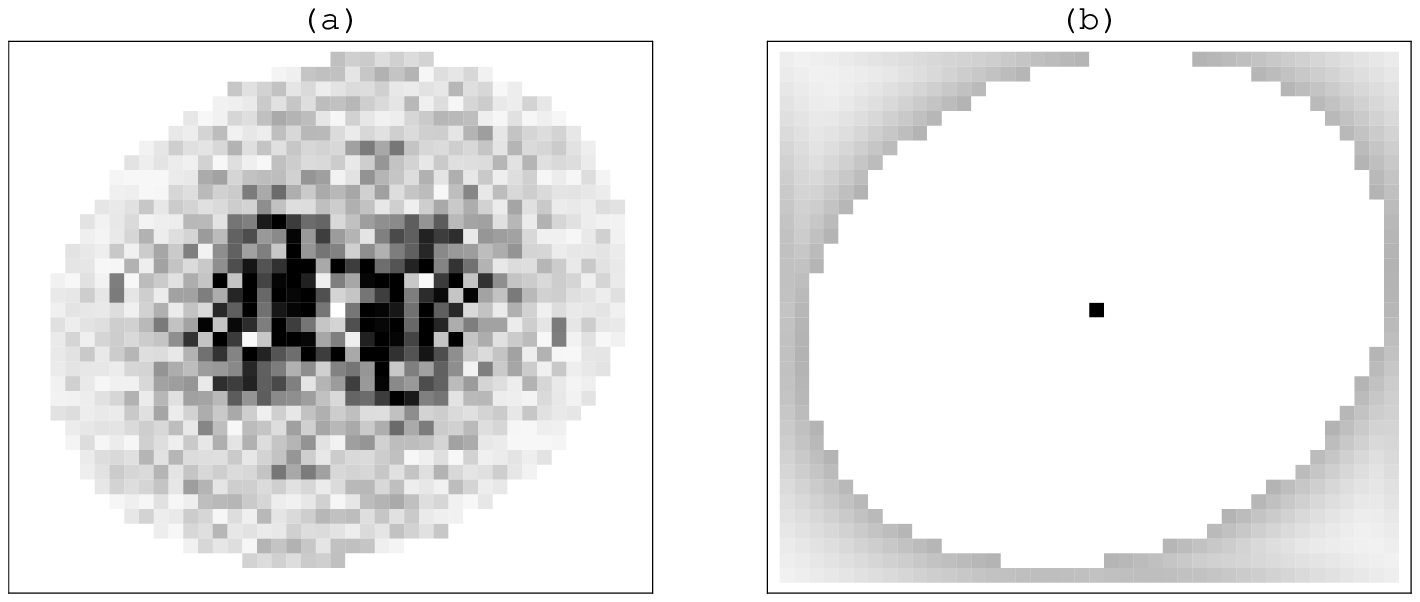}
\end{figure}

\newpage

\begin{figure}
\caption{Atomicity projections of a two dimensional density.
(a) Density $\rho$ within one unit cell.
(b) $\Pi_{\mathrm{A+}}(\rho)$, the density closest to (a) having five positive atoms.
(c) $\Pi_{\mathrm{A}}(\rho)$, same as (b) but with the signs of the atoms unspecified.}
\includegraphics{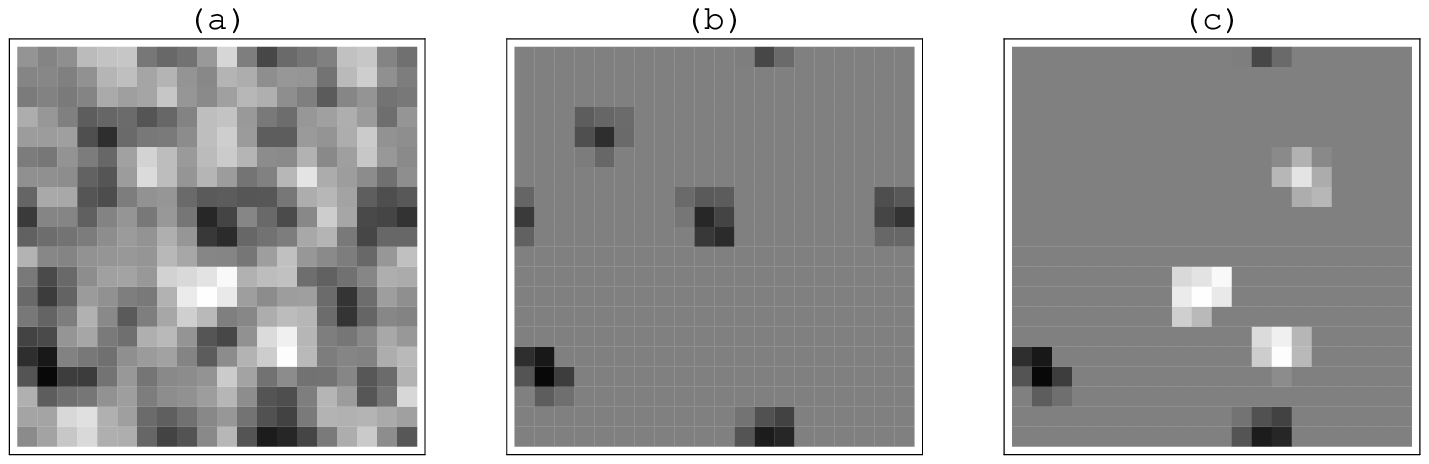}
\end{figure}

\begin{figure}
\caption{Average number of difference map iterations required to solve a 148-atom structure (Table 1, ref. 5)
for the parameter range $0.4\le\beta\le0.8$ (over 200 solutions per point).}
\includegraphics{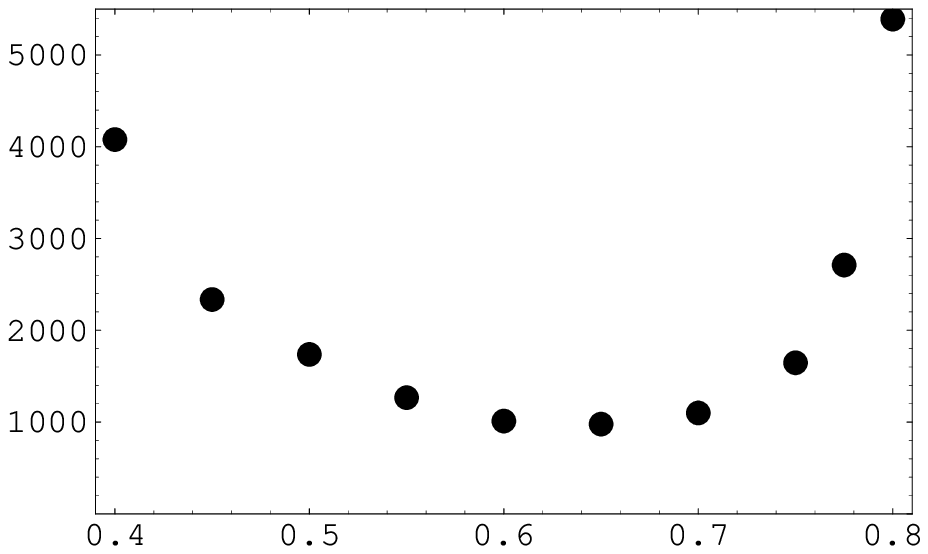}
\end{figure}

\newpage

\begin{figure}
\caption{Evolution of the error estimate $\epsilon$ with difference map iteration for
two 148-atom data sets: (top) synthetic data for ideal $3\times 3\times 3$ atoms and no H-atoms,
(bottom) experimental data (Table 1, ref. 5).}
\includegraphics{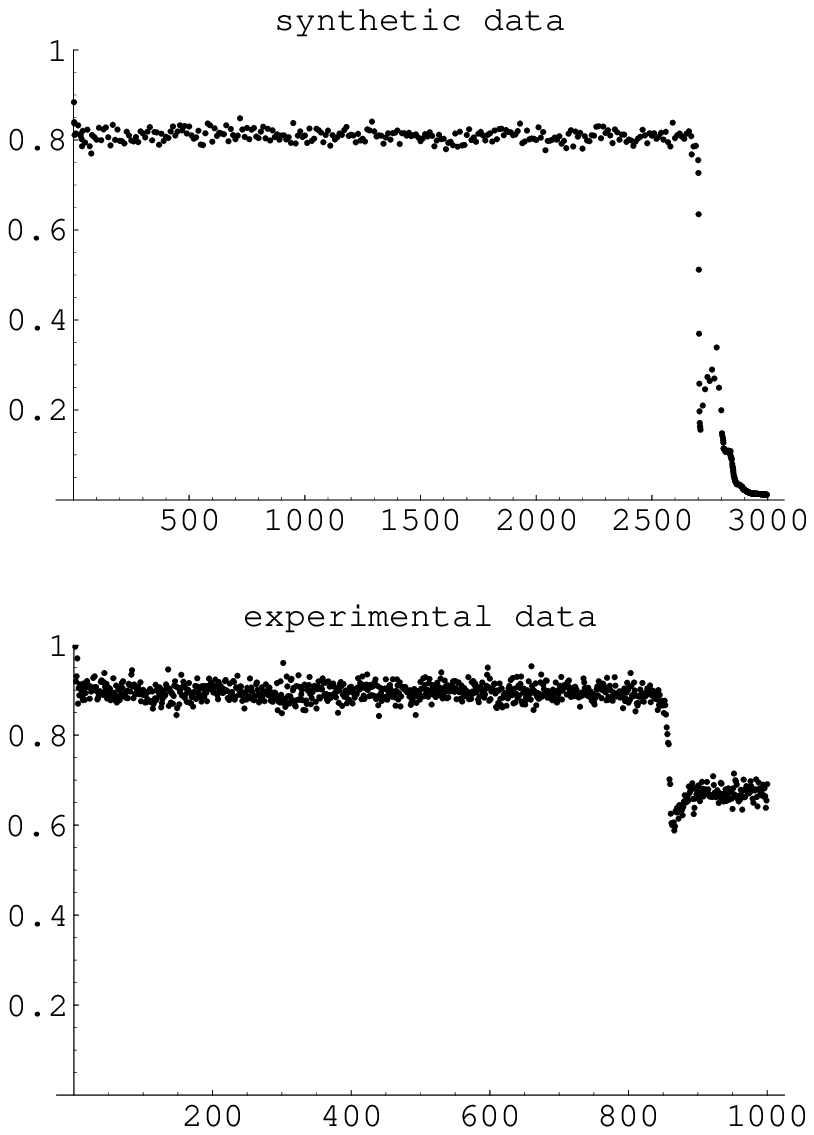}
\end{figure}

\newpage

\begin{figure}
\caption{Detailed behavior of the densities in the plane $x=0$ for the solution shown at the top of Fig. 5:
(a-b) consecutive densities $\rho(100)$ and $\rho(101)$,
(c) fixed point density $\rho^*\approx\rho(3000)$,
(d) solution density $\rho_{\mathrm{sol}}=\Pi_{\mathrm{F}}\circ f_{\mathrm{A+}}(\rho^*)$.}
\includegraphics{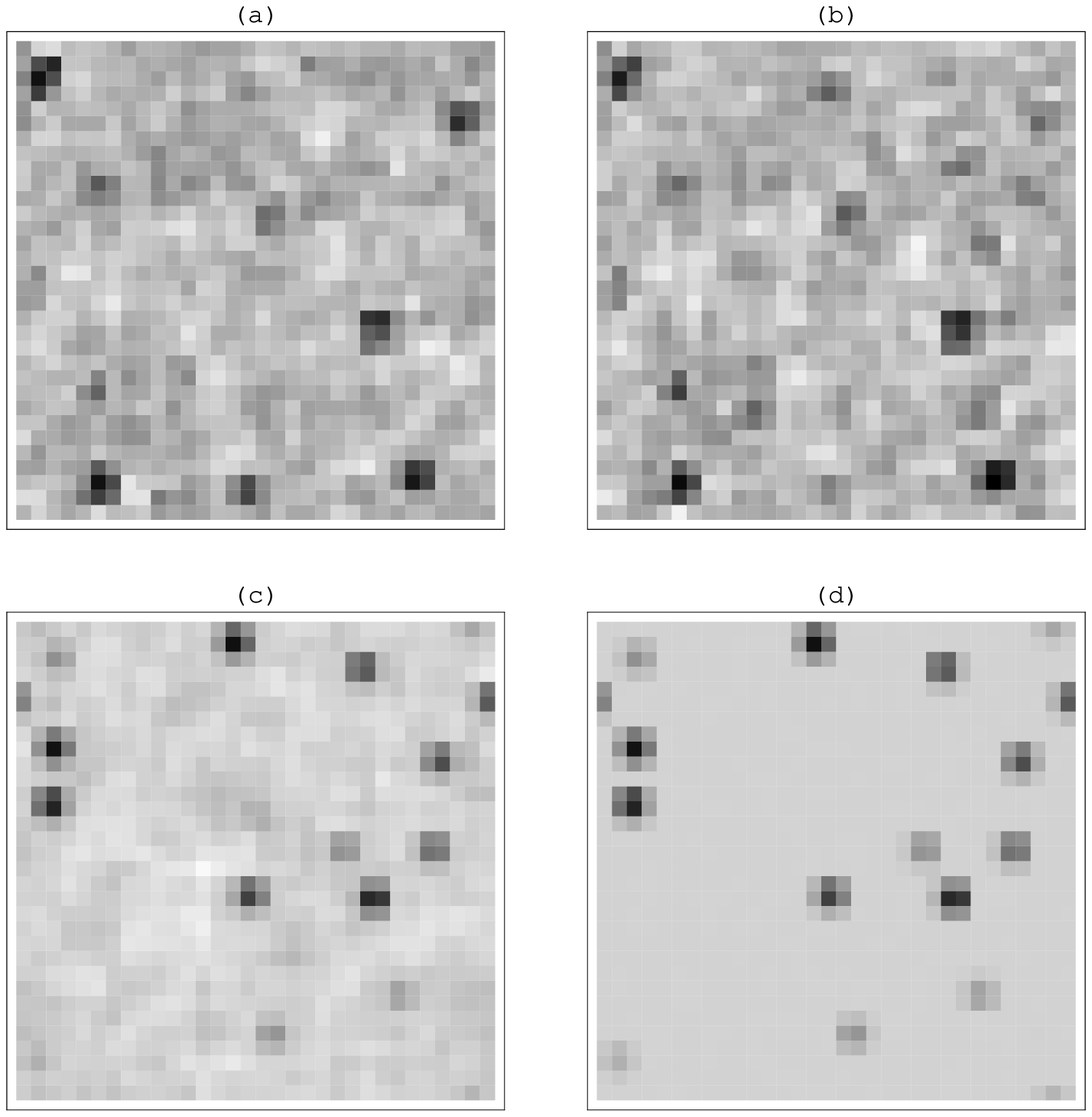}
\end{figure}

\newpage

\begin{figure}
\caption{Distribution of difference map solution times (iterations) for the 148-atom peptide structure 
and
$\beta=0.7$. Solution times on the abscissa are normalized by the mean number of iterations, 1100. The
curve shows the exponential distribution with mean unity, predicted by the strongly mixing hypothesis
of difference map dynamics.}
\includegraphics{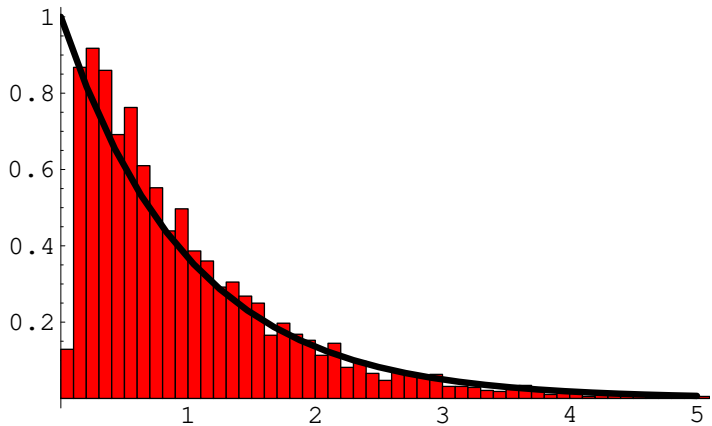}
\end{figure}

\begin{figure}
\caption{\texttt{atom\_retriever} solution for the neutron data set of the mineral montebrasite (Table 1); 
(left) near-monotonic decrease of the error with iteration, (right) scattering density
$\rho_{\mathrm{sol}}$ in a plane of the
structure showing an atom (Li) with negative scattering length (light contrast).}
\includegraphics{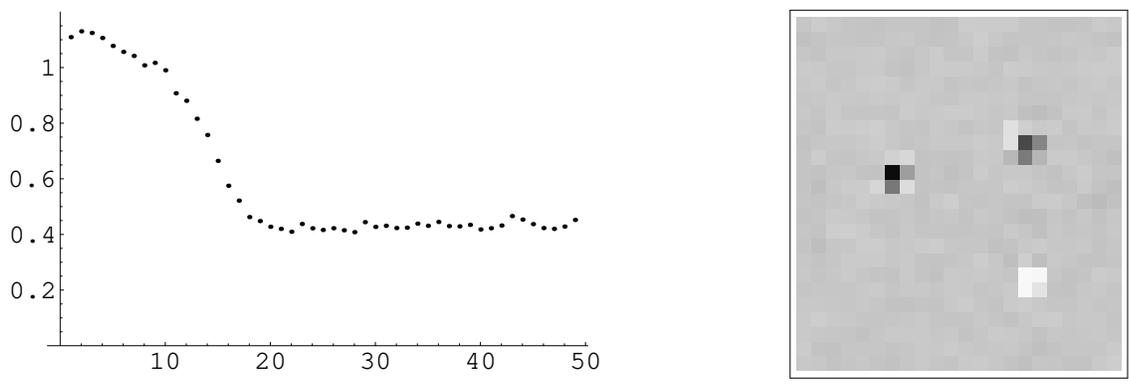}
\end{figure}

\newpage

\begin{figure}
\caption{\texttt{atom\_retriever} solution of the synthetic $\alpha$-helical bundle (Table 1); 
(top) evolution of the error estimate, (bottom) electron density contours in a plane containing
the chlorine ion (top, left of center).}
\centerline{\scalebox{1.}{\includegraphics{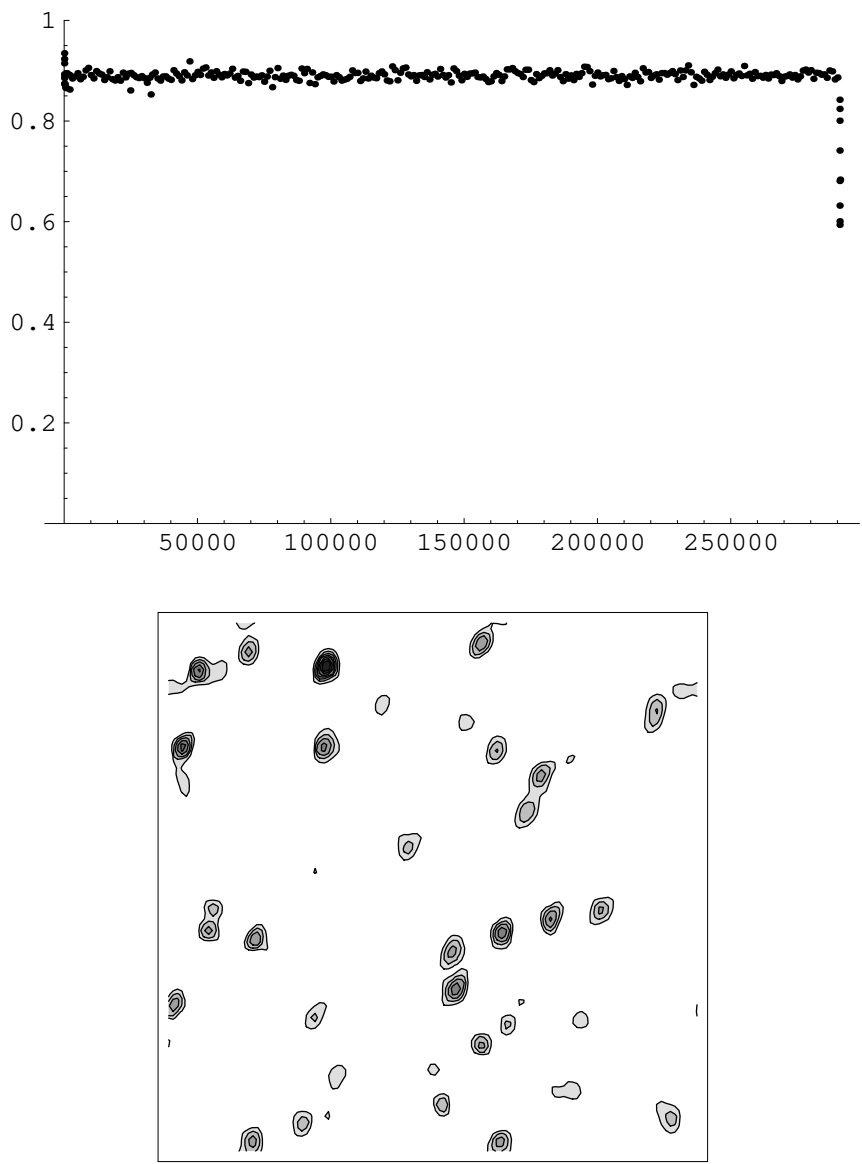}}}
\end{figure}

\end{document}